\documentclass{elsart}
\usepackage{graphicx}
\usepackage{epsfig}
\usepackage{amssymb}
\journal{Chemical Physics Letters}

\begin{document}

\begin{frontmatter}

\title{Complex reaction noise in a molecular quasispecies model}

\author[CAB]{David Hochberg\corauthref{cor}},
\corauth[cor]{Corresponding author.} \ead{hochberg@laeff.esa.es}
\author[CAB]{Mar\'{i}a-Paz Zorzano},
\ead{zorzanomm@inta.es}
\author[CAB,UCM]{Federico Mor\'{a}n}
\ead{fmoran@bio.ucm.es}

\address[CAB]{Centro de Astrobiolog\'{\i}a (CSIC-INTA), Ctra. Ajalvir
Km. 4, 28850 Torrej\'{o}n de Ardoz, Madrid, Spain}
\address[UCM]{Departamento de Bioqu\'{\i}mica y
Biolog\'{\i}a Molecular, Facultad de Ciencias Qu\'{\i}micas,
Universidad Complutense de Madrid, Spain}
\begin{abstract}
We have derived exact Langevin equations for a model of quasispecies
dynamics. The inherent multiplicative reaction noise is
\textit{complex} and its statistical properties are specified
completely. The numerical simulation of the complex Langevin
equations is carried out using the Cholesky decomposition for the
noise covariance matrix. This internal noise, which is due to
diffusion-limited reactions, produces unavoidable spatio-temporal
density fluctuations about the mean field value. In two dimensions,
this noise strictly vanishes only in the perfectly mixed limit, a
situation difficult to attain in practice.
\end{abstract}

\end{frontmatter}

\section{Introduction}
The study of replicator models of prebiotic evolution has received
considerable theoretical attention during the past three decades
\cite{ES}. There is a renewed and active interest in this subject
owing to the crucial fact that the dynamics of real viral
populations is known to be described by quasispecies
\cite{Domingo,BE1,BE2}. To date, the bulk of the theoretical work
devoted to the analysis of quasispecies dynamics has tacitly assumed
spatially homogeneous conditions \cite{ES,Domingo,BE1,BE2,NAMM}. The
importance of diffusive forces has begun to be recognized and taken
into account \cite{CN,CB}, but rather less attention, if any, has
been payed to the presence of the unavoidable internal density
fluctuations \cite{HZM} that are necessarily present in all
realistic incompletely mixed and diffusing systems of reacting
agents \cite{Epstein}. In the deterministic approach to chemical and
molecular species that diffuse and react, the fluctuations are
simply ignored. Nevertheless, it is well known that if the spatial
dimensionality $d$ of the system is smaller than a certain upper
critical dimension $d_c$, the intrinsic fluctuations do play a
decisive role in the asymptotic late time behavior of decay rates,
and the results obtained from the mean field equations are not
correct \cite{KangRedner}. Even far from asymptotia, these
fluctuations can strongly affect the dynamics on local spatial and
temporal scales \cite{ZHM}. The mean field limit is strictly valid
only in the infinite diffusion limit, because the reactions
themselves induce local microscopic density fluctuations that can be
amplified by the underlying nonlinear dynamics. This internal noise
is multiplicative, and scales as the square root of the product of
concentrations, with the consequence that the noise does not
necessarily vanish in the large particle-number limit.

Given the relevance of molecular replicator models such as the
quasispecies for viral dynamics and the key role played by
fluctuations in nonlinear systems, this Letter has a twofold
objective. First, we outline a rigorous derivation of the exact
stochastic partial differential equations (SPDE) that govern the
dynamics of a simple quasispecies model (denotes in this work a
system comprised of a master plus mutant molecular species in
competition for limited resources). Second, we study the nature of
the internal fluctuations on the replicator evolution. But in order
to achieve this latter goal, we must also develop and apply a method
suitable for simulating multi-component SPDE's with \textit{complex}
noise \cite{DFHK}. The validity of this approach is confirmed by
numerical simulation. Neither of these objectives has been achieved
before. We illustrate this method with a simple model that is
analytically tractable, but the technique given here can be applied
to more realistic chemical and molecular reaction processes.

\section{Model}
We consider a molecular quasispecies model with error introduced via
the faulty self-replication of the master copy into a mutant
species. The mutant species or, error-tail, undergoes non-catalyzed
self-reproduction, but has no effect on the master species. The
system is closed, only energy can be exchanged with the
surroundings, where activated monomers react to build up
self-replicative units. These energy rich monomers are regenerated
from the by-product of the reactions by means of a recycle mechanism
(driven by an external source of energy) maintaining the system out
of equilibrium. The closure of the system directly imposes a
selection pressure on the population. In what follows, ${M}^*,I,I_e$
denote the concentrations of the activated energy rich monomers, the
master and the mutant copies, respectively. The kinetic constants
are introduced in the following reaction scheme.

Accurate replication with rate $AQ$:
\begin{equation}\label{ancatrep}
\textrm{M}^* + \textrm{I} \stackrel{AQ}{\longrightarrow}
2\textrm{I}.
\end{equation}
Error replication with rate $A(1-Q)$:
\begin{equation}\label{encatrep}
\textrm{M}^* + \textrm{I} \stackrel{A(1-Q)}{\longrightarrow}
\textrm{I} + \textrm{I}_e.
\end{equation}
Mutant-species replication with rate $A_e$:
\begin{equation}\label{erep}
\textrm{M}^* + \textrm{I}_e \stackrel{A_e}{\longrightarrow}
2\textrm{I}_e.
\end{equation}
Degradation of the master and mutant copies into monomers with rates
$r,r_e$, respectively:
\begin{equation}\label{degradI}
\textrm{I} \stackrel{r}{\longrightarrow} \textrm{M}, \qquad
\textrm{I}_e \stackrel{r_e}{\longrightarrow} \textrm{M}.
\end{equation}
Re-activation of energy-depleted monomers (induced by a generic
energy-driven reaction):
\begin{equation}\label{activation}
\textrm{M} \stackrel{energy}{\longrightarrow} \textrm{M}^*.
\end{equation}
The quality factor $Q \in [0,1]$. In order to keep the following
development mathematically manageable, we will assume that the
monomer reactivation step Eq.(\ref{activation}) proceeds
sufficiently rapidly, so that we can in effect, regard the decay of
$I$ and $I_e$, Eq.(\ref{degradI}), plus the subsequent reactivation
$\textrm{M} \stackrel{energy}{\longrightarrow} \textrm{M}^*$ as
occurring in \emph{one} single step, i.e.,
\begin{eqnarray}
\label{degradIprime} \textrm{I} \stackrel{r}{\longrightarrow}
\textrm{M} \stackrel{energy}{\longrightarrow} \textrm{M}^*
&\Rightarrow& \textrm{I} \stackrel{r'}{\longrightarrow} \textrm{M}^* \\
\label{degradIeprime} \textrm{I}_e \stackrel{r_e}{\longrightarrow}
\textrm{M} \stackrel{energy}{\longrightarrow} \textrm{M}^*
&\Rightarrow& \textrm{I}_e \stackrel{r'_e}{\longrightarrow}
\textrm{M}^* .
\end{eqnarray}
Here, $r'$, $r_e'$ are correspondingly, the combined
\textit{effective} decay plus regeneration rates of $I,I_e$ to the
reactivated monomers $M^*$, although in what follows, we will simply
write $r$ and $r_e$. This means that our continuum field theory,
will depend on three $(M^*,I,I_e)$, instead of four $(M,M^*,I,I_e)$
concentration field variables. If we suppose the system is being
bathed continuously by an external energy source, the monomer
reactivation Eq.(\ref{activation}) is occurring continuously, and
this should be a reasonable approximation.  To complete the
specification of the model, we allow for spatial diffusion. This is
incorporated into the master equation associated with the above
reaction scheme. The diffusion constants are denoted by $D_s,D_I$
and $D_e$ for $M,I$ and $I_e$, respectively. The constraint of
constant total particle number is \textit{automatically} satisfied
by the continuous chemical concentration fields in the mean-field
limit. Most importantly, this provides a selection pressure on the
quasispecies.

\section{Complex Langevin equations}
It is straightforward to write down the continuous time master
equation corresponding to the above reaction scheme  Eqs.
(\ref{ancatrep},\ref{encatrep},\ref{erep},\ref{degradIprime},\ref{degradIeprime}).
This master equation is then mapped to a second-quantized
description following Doi \cite{Doi}. From Doi's operator language,
we then pass to a path integral representation in terms of
continuous stochastic fields \cite{Peliti} to obtain the following
action $S$ valid for any space dimension $d$ (D. Hochberg, M.-P.
Zorzano and F. Mor\'{a}n, unpublished notes):
\begin{eqnarray}\label{shiftaction}
S &=& \int dt d^dx \, \Big(\tilde a\big(\partial_t a -D_s\nabla^2a +
A ab + A_eac-rb-r_ec\big)  \nonumber \\
&+& \tilde b\big(\partial_tb - D_I\nabla^2b -AQ ab +
rb\big)\nonumber \\ &+& \tilde c\big(\partial_tc - D_e\nabla^2c
-A(1-Q)ab -A_e ac + r_ec
\big)  \nonumber \\
&-&  AQab\big({\tilde b}^2 - \tilde a\tilde b \big)
-A(1-Q)ab\big(\tilde b \tilde c -  \tilde a \tilde b\big) - A_e ac
\big({\tilde c}^2 -\tilde a \tilde c \big) \Big).
\end{eqnarray}
In the \textit{mean field} limit the continuous field variables
$a,b,c$ in Eq.(\ref{shiftaction}) correspond to the physical
densities of the molecules $M,I,I_e$, respectively. The fields
$\tilde a,\tilde b,\tilde c$, are conjugate to the internal
fluctuations. Only if this noise is real, do $a,b,c$ continue to
represent the physical densities. Otherwise, for imaginary or
complex noise, these fields will likewise be imaginary or complex,
but their stochastic averages $\langle a \rangle, \langle b\rangle$
, and $\langle c \rangle$ do however correspond to the real physical
densities \cite{HT}.

Since the $\tilde a,\tilde b,\tilde c$ fields appear quadratically
in $S$, we can make use of the Hubbard-Stratanovich transformation
to integrate over them exactly in the path integral $\int \mathcal{
D}a \mathcal{D}{\tilde a} \mathcal{D}b \mathcal{D}{\tilde b}
\mathcal{D}c \mathcal{D}{\tilde c} \, e^{-S[a,\tilde a, b, \tilde b,
c, \tilde c]}$. This final step yields a product of delta-functional
constraints which directly imply a set of exact coupled stochastic
partial differential equations with specific noise properties (see
below).

In particular, in a two-dimensional space and making the reasonable
assumption that both master and mutant species diffuse equally
$D_{Ie}=D_I=D$ and have equal effective degradation plus
reactivation rates, i.e. $r_e/r=1$, the reaction-diffusion system in
\textit{dimensionless} variables \footnote{Define the dimensionless
fields: $\bar a = \frac{A}{r}a, \, \bar b = \frac{A}{r}b$, and $\bar
c = \frac{A_e}{r}c$; the dimensionless time $\tau = rt$ and spatial
coordinates $\hat x_j = \Big(\frac{r}{D_I}\Big)^{1/2}\,x_j$. Their
corresponding derivative operators are $\frac{\partial}{\partial
\tau} = \frac{1}{r}\frac{\partial}{\partial t}$, and ${\hat
\nabla}^2 = \big(\frac{D_I}{r}\big)\nabla^2.$ The dimensionless
noises are defined by ${\hat \eta_a} = \frac{A}{r^2}\eta_a, {\hat
\eta_b} = \frac{A}{r^2}\eta_b$, and ${\hat \eta_c} =
\frac{A_e}{r^2}\eta_c$. } is given as follows:
\begin{eqnarray}\label{QEN}\nonumber
\frac{\partial \bar{a}}{\partial \tau}&=&(D_s/D) \nabla ^{2} \bar{a}
- \bar{a}\bar{b}-\bar{a}\bar{c}+\bar{b}+ \frac{1}{\delta}\bar{c} +
\hat{\eta}_a\\ \nonumber \frac{\partial \bar{b}}{\partial
\tau}&=&\nabla ^{2} \bar{b}  +Q \bar{a}\bar{b}- \bar{b} +
\hat{\eta}_b \\ \nonumber \frac{\partial \bar{c}}{\partial
\tau}&=&\nabla ^{2} \bar{c}  + \delta (1-Q)\bar{a}\bar{b}+
\delta\bar{a}\bar{c} - \bar{c}+ \hat{\eta}_c,
\end{eqnarray}
where $\nabla ^{2}=\frac{\partial ^{2}}{\partial \hat
x^{2}}+\frac{\partial ^{2}}{\partial \hat y^{2}}$ and
$\vec{{\eta}}=(\hat{\eta} _b,\hat{\eta} _c,\hat{\eta} _a)$ is the
noise with $\langle\vec{{\eta}}\rangle=0$ and correlation matrix
$B=\langle\vec{{\eta}}\vec{{\eta}}'^T\rangle$:
\begin{eqnarray}\nonumber
B(\hat{x},\hat{y};\tau)=\left(
\begin{array}{ccc}
\epsilon Q \bar{a} \bar{b} & \frac{1}{2}\epsilon\delta (1-Q)
\bar{a} \bar{b} &  -\frac{1}{2}\epsilon \bar{a} \bar{b} \\
\frac{1}{2}\epsilon\delta (1-Q)\bar{a} \bar{b} & \epsilon \delta^2
\bar{a} \bar{c} &  -\frac{1}{2}\epsilon \delta \bar{a} \bar{c}\\
-\frac{1}{2}\epsilon \bar{a} \bar{b} & -\frac{1}{2}\epsilon \delta
\bar{a} \bar{c} & 0
\end{array}
\right).
\end{eqnarray}
The initial condition $\bar{a}_0,\bar{b}_0,\bar{c}_0$ and the ratio
of replication rates $\delta = A_e/A < 1$ defines the  {\em scaled}
number of particles in the closed system $\bar{N}=\int \int
d\hat{x}d\hat{y}(\bar{a}_0+\bar{b}_0+\frac{\bar{c}_0}{\delta})$. In
the two-dimensional case the total number of particles is given by
the ratio $N=\bar{N}/\epsilon$ where $\epsilon=A/D_I$ is the ratio
of the reaction to the diffusion processes (in any dimension $d$, we
have $\epsilon=(r/D_I)^{d/2}A/r$). In the absence of noise
$\vec{{\eta}}$ this reaction-diffusion system has a set of
homogeneous solutions: (i) $\bar{b}=\bar{c}=0,\ \bar{a}=\bar{N}$,
(the trivial solution); (ii) $\bar{b}=0,\
\bar{c}/\delta=\bar{N}-\frac{1}{\delta},\ \bar{a}=\frac{1}{\delta}$,
if $\delta>1/\bar{N}$ and $\delta\leq Q$; and (iii)
$\bar{c}/\delta=\frac{(\bar{N}Q-1)(1-Q)}{Q(1-\delta)},\
\bar{b}=\frac{(\bar{N}Q-1)(Q-\delta)}{Q(1-\delta)}$, $\bar{a}=1/Q$
if $Q>\delta$ and $Q>1/\bar{N}$. We want to understand how this
deterministic, mean-field homogeneous solution, is modified by the
internal multiplicative noise term with defined spatio-temporally
varying covariance (the fields vary in both space and time). From
inspection of the covariance matrix $B$ we see that this noise is
proportional to $\epsilon$, the ratio between the reaction and
diffusion processes. In the limit $\epsilon\rightarrow 0$ (infinite
diffusion, i.e., a perfectly homogeneous system) there are no
fluctuations, and we recover the mean field result. The noise term
vanishes when $\bar{b}(\hat x,t)=\bar{c}(\hat x,t)=0$ (the trivial
solution is an inactive state) and/or when $\bar{a}(\hat x,t)=0$. If
the system stays close to the mean field result, then $\bar{a}=1/Q$
and $\bar{b}$ and $\bar{c}$ scale as $(\bar{N}Q-1)$. Therefore the
noise covariance is expected to scale with $\epsilon\bar{N}$.  Our
main interest here is the limit when the stochastic term makes a
significant contribution $\epsilon\bar{N}=\epsilon^2 N>1$. In this
regime the problem can not be analyzed by perturbation theory and
thus must be treated numerically \footnote{After the space and time
discretization of the stochastic partial differential equations, the
numerical integration of the finite-difference equations has been
performed using forward Euler with time step $\Delta \tau=5\times
10^{-4}$ and a spatial mesh of $154\times 154$ cells with cell size
$\Delta \hat{x}=\Delta \hat{y}= 0.35$ and periodic boundary
conditions.}.

Note that $B$ is a symmetric matrix with $\det
B=-\frac{Q}{4}\bar{a}^3 \bar{b} \bar{c} \epsilon ^3 \delta
^2(\bar{b} +\bar{c} )$, i.e. it is negative definite. Thus it either
has one or three negative eigenvalues, suggesting that at least one
noise component will have negative auto-correlation, and will
therefore be imaginary. We expect this situation to be relatively
common in reaction-diffusion systems with multiple species.

\section{Results}
We next apply the Cholesky decomposition, which is used when the
symmetric matrix is positive definite, to relate this correlated
noise to an uncorrelated one. We apply this algorithm for the first
time here to a case with negative definite correlation matrix to
obtain the ``square root'' $L$, where some of the terms are
manifestly imaginary. This is very useful for relating the noise
$\vec{\eta}$ to a new {\em real} white Gaussian noise $\vec{\xi}$
with $\langle\vec{\xi}\vec{\xi}'^T\rangle=1$, such that
$\vec{\eta}=L\vec{\xi}$ \footnote{ For an $M\times M$ symmetric
matrix $B$, we can apply the Cholesky Decomposition $B=L L^T$ to
extract the square root of the matrix in the form of a lower
triangular matrix $L$ with
$L_{ii}=\sqrt{B_{ii}-\sum_{k=1}^{i-1}L^2_{ik}}$ and
$L_{ji}=\frac{1}{L_{ii}}(B_{ji}-\sum_ {k=1}^{i-1}L_{ik}L_{jk})$ for
$j=i+1,...,M$. Note that $\langle\vec{\eta}\vec{\eta}'^T\rangle=
\langle L\vec{\xi}\vec{\xi}'^TL^T \rangle=L\langle
\vec{\xi}\vec{\xi}'^T \rangle L^T=L L^T=B$ is automatically
satisfied. }:
\begin{eqnarray}
\hat{\eta_{b}}=&\sqrt{\epsilon \bar{a}}& \sqrt{Q\bar{b}}\xi_{1},
\qquad \hat{\eta_{c}}= \sqrt{\epsilon \bar{a}}
\left(\frac{\delta(1-Q)
\sqrt{\bar{b}}}{2\sqrt{Q}}\xi_{1}+\frac{\delta
\sqrt{4Q\bar{c}-(1-Q)^2 \bar{b}}}{2\sqrt{Q} }\xi_{2}\right),
\nonumber \\
\label{eta_a} \hat{\eta_{a}}=&\sqrt{\epsilon \bar{a}}&
\left(-\frac{\sqrt{b}}{2\sqrt{Q} }\xi_{1}+\frac{1}{2\sqrt{Q} }
\frac{(1-Q)\bar{b}-2 Q\bar{c}}{\sqrt{4Q\bar{c}-(1-Q)^2
\bar{b}}}\xi_{2}\right)\nonumber \\ + &\sqrt{\epsilon \bar{a}}&
\frac{\sqrt{-1}\sqrt{Q}\sqrt{ \bar{b}\bar{c}+\bar{c}^2}}{\sqrt{
4Q\bar{c}-(1-Q)^2 \bar{b}}}\xi_{3}.
\end{eqnarray}
This transformation allows us to separate the real and imaginary
parts of the noise, a extremely useful feature to have for setting
up and running a stable numerical simulation.

The noise $\hat{\eta_{a}}$ for the nutrient field $\bar{a}$,
Eq.(\ref{eta_a}), {\em always} has an imaginary component and since
all the equations are coupled to the nutrient $\bar{a}$, even if the
initial condition for $\bar{a},\bar{b},\bar{c}$ is real, the fields
will, in principle, have both real and imaginary parts, and thus we
have to solve a system of 6 partial differential equations, one for
each field ($\Re \bar{a},\Im\bar{a},\Re \bar{b},\Im\bar{b},\Re
\bar{c},\Im\bar{c}$) with two-dimensional diffusion and noise. We
expect the imaginary fields to be zero in the average, since the
stochastic averages $\langle a \rangle, \langle b\rangle$ , and
$\langle c \rangle$ correspond to the physical densities \cite{HT}.

\begin{figure}[h]
\begin{center}
\begin{tabular}{ccc}
$\bar{a}(\hat{x},\hat{y},\tau=25)$ & $\bar{b}(\hat{x},\hat{y},
\tau=25)$ & $\bar{c}(\hat{x},\hat{y},\tau=25)$ \\
\includegraphics[width=0.30 \textwidth]{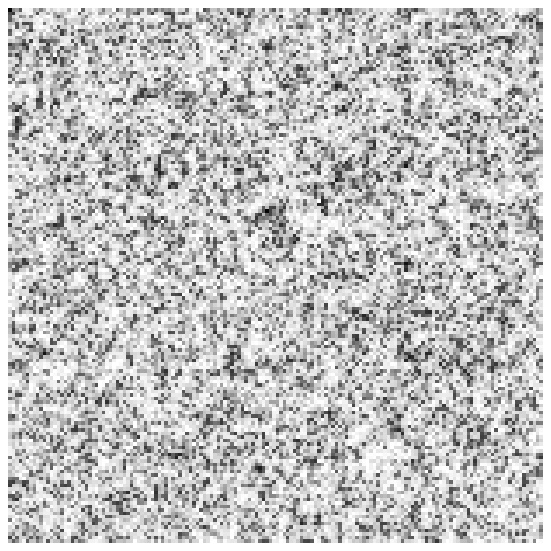} &
\includegraphics[width=0.30  \textwidth]{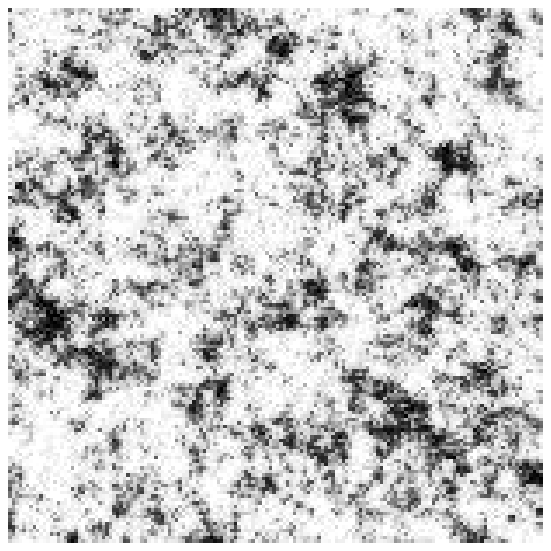} &
\includegraphics[width=0.30 \textwidth]{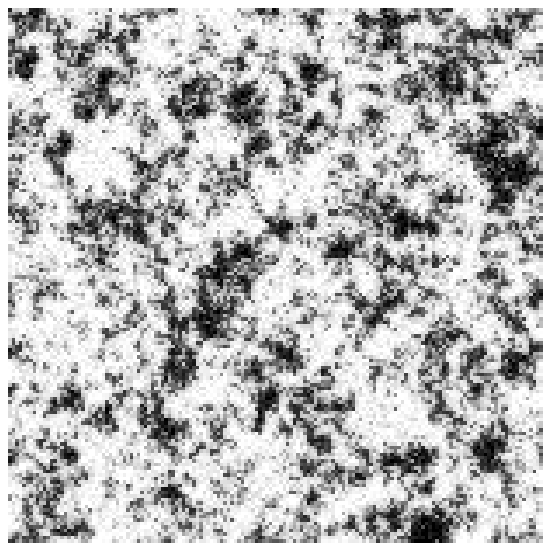}\\
\end{tabular}
\caption{\label{fig}Long-time spatial distribution of the fields
when $Q=0.92,\delta=0.8,D_s/D=10$ and  $\epsilon=1$,
$N=\bar{N}=10^{3}$. The three fields show
 fluctuations greater than $1\%$ about the mean-field value.
}
\end{center}
\end{figure}

\begin{figure}[h]
\begin{center}
\begin{tabular}{cc}
\includegraphics[width=0.47 \textwidth]{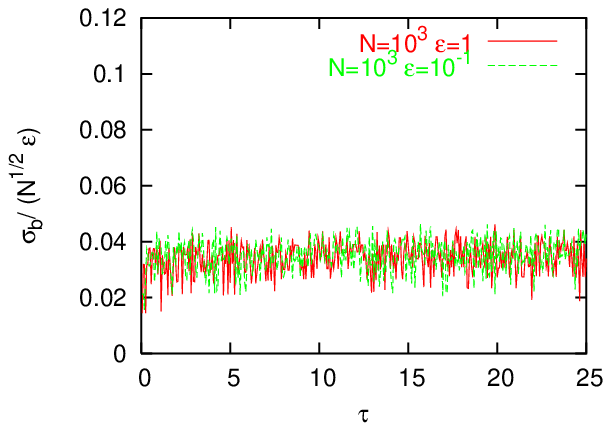}
&
\includegraphics[width=0.48 \textwidth]{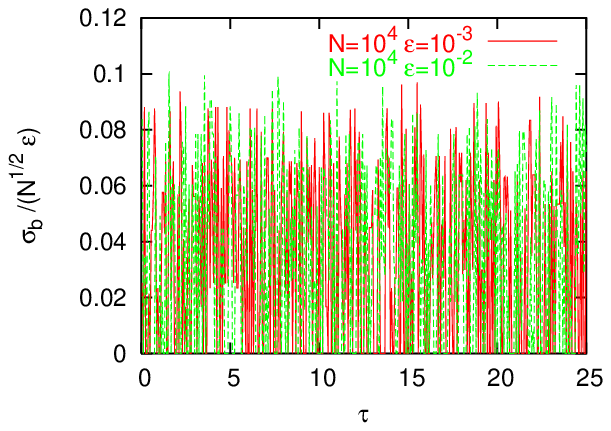}\\
\end{tabular}
\caption{\label{fig2}Time evolution of the fluctuations $\sigma
_{\bar{b}(\tau)}=\sqrt{\langle(\bar{b} (\tau)-\langle\bar{b}
(\tau)\rangle)^{2}\rangle}$ scaled by
$\sqrt{N}\epsilon=\sqrt{\bar{N}\epsilon}$ for $N=10^3$ (left) and
$N=10^{4}$ (right) with $\epsilon$ between $10^{-3}$ and $1$. The
fluctuations around the mean-field value increase with $N$ and
$\epsilon$.}
\end{center}
\end{figure}

Integrating the system numerically we confirmed that the scaled
number of particles was conserved and remained {\em real}
$\Re\bar{N}=\bar{N}$, and $\Im\bar{N}=0$ (within computational
errors)  \footnote{In fact, for computational purposes, $\Im\bar{N}$
was set as small as possible. We confirmed that both $\Re
\bar{N}=\int \int d\hat{x}d\hat{y}(\Re \bar{a}+
\Re\bar{b}+\frac{\Re\bar{c}_0}{\delta})=\bar{N}$ and $\Im
\bar{N}=\int \int d\hat{x}d\hat{y}(\Im\bar{a}+
\Im\bar{b}+\frac{\Im\bar{c}_0}{\delta})=\bar{N}\times10 ^{-7}$ are
conserved during the integrating time and that the results were
independent of $\Im \bar{N}$.}. We found that, independently of the
initial condition and noise intensity, it always converged to the
same state with spatial fluctuations around this value: $\langle
\frac{\bar{c}}{\delta}\rangle_{\hat{x},\hat{y}}=\frac{(\bar{N}Q-1)(1-Q)}{Q(1-\delta)},\
\langle \bar{b}
\rangle_{\hat{x},\hat{y}}=\frac{(\bar{N}Q-1)(Q-\delta)}{Q(1-\delta)}$,
$ \langle \bar{a} \rangle_{\hat{x},\hat{y}}=1/Q$,  which is the
mean-field solution \footnote{The imaginary part converged to $
\langle \Im\bar{c}/\delta \rangle_{\hat{x},\hat{y}}=\frac{\Im
\bar{N}(1-Q)}{1-\delta},\ \langle \Im\bar{b}
\rangle_{\hat{x},\hat{y}}=\frac{\Im \bar{N}(Q-\delta)}{1-\delta}$,
$\langle \Im \bar{a}\rangle_{\hat{x},\hat{y}}=0$. The spatial
average is denoted by $\langle \bullet \rangle_{\hat{x},\hat{y}}$,
which by the ergodic hypothesis, is equivalent to averaging over the
noise.} An example of these spatial fluctuations is shown in Fig.
(\ref{fig}) at $\tau=25$ for $\epsilon=1$ and $N=\bar{N}=10^{3}$,
with $D_s/D=10$, $Q=0.92$, and $\delta=0.8$. In the case of
$\bar{b}$ and $\bar{c}$, the densities show spatial fluctuations of
up to $+1\%$ (in white) and $-1\%$ (in black) with respect to the
mean field value. Inspection of the figure clearly shows that the
master copy and mutant species are spatially anti-correlated. In the
case of $\bar{a}$, the fluctuations range in the order of $\pm 5\%$
about the mean field value $\bar{a}=1/Q$.
Next we explore the dependence of the spatio-temporal field
fluctuations $\sigma ^{2}_{\bar{b} (\tau)}=\langle (\bar{b}
(\tau)-\langle \bar{b}
(\tau)\rangle_{\hat{x},\hat{y}})^{2}\rangle_{\hat{x},\hat{y}}$ etc.,
in $\epsilon$ and $N$. In Fig. \ref{fig2}-left we show the deviation
$\sigma_{\bar{b}(\tau)}$  scaled by $\sqrt{N}\epsilon$ for
$N=\frac{\bar{N}}{\epsilon}=10^3$ and two different cases of
$\epsilon$. As expected, $\sigma _{\bar{b}}$ scales with
$\sqrt{N}\epsilon$, in this case as $0.04\times \sqrt{N}\epsilon$.
Next we show the same for $N=10^{4}$, the mean value of $\sigma
_{\bar{b}}$ is the same but the deviations with respect to it (which
is related to the kurtosis of the distribution) increase with $N$.
Notice that the \textit{relative} weight of these fluctuations with
respect to the mean-field value decreases with the number of
particles since $\bar{b}\propto \bar{N}$  and therefore
$\sigma_{\bar{b}}/\bar{b} \propto 0.04 \sqrt{N}\epsilon/(N\epsilon)
\propto 0.04/\sqrt{N}$.

\section{Conclusions}
Deriving the Langevin equation description from the Master equation
provides a quick analytic survey of the global system dynamics
(stationary states, etc), as well as the strength and
characteristics of the internal noise and its dependence on the
spatial dimensionality. For any dimension, the system can be solved
numerically with high accuracy. This is in contrast to the direct
Monte Carlo simulation of the system, such as in [16], where no
preliminary analytic information can be given on the effects of
reaction noise or dimensionality dependence. We have shown that the
local deviations with respect to the mean-field solution scale with
the number of particles and the ratio between the reaction and
diffusion processes. The fluctuations vanish only in the perfectly
mixed limit, when the diffusion processes are infinitely fast
($\epsilon=0$). Thus, even in the case of very high diffusion rates,
an autocatalytic system can show important deviations with respect
to the perfect mixing limit, provided that the number of particles
in the reactor system is sufficiently high. For systems with a
higher degree of non-linearity (i.e., third-order), these deviations
may eventually lead the system to new asymptotic states and also
induce the formation of spatial patterns \cite{HZM}. Regarding the
error catastrophe, this concept differs from virology to molecular
replicator dynamics. In the former, the error catastrophe implies
the proliferation of lethal mutations with the subsequent loss of
viral infectivity, and consequently, the extinction of the viral
population. This is reflected in our model as the trivial solution
(i). On the other hand, solution (ii) corresponds to the error
catastrophe state predicted by molecular replicator dynamics where
there is no wild type population, while there is an error tail. We
have found that this solution is an absorbing barrier. Finally,
spatial diffusion and internal noise must in fact play an important
role in the correct description of viral infection dynamics, where
typically, bottlenecks of different intensities may lead to
infection of host cells by limited numbers of viral particles
\cite{ELM}.

We thank Carlos Escudero for many useful discussions and for working
through some preliminary analytic calculations and Esteban Domingo
for reading the manuscript and suggesting additional references.
M.-P.Z. is supported by an INTA fellowship. The research of D.H. is
supported in part by funds from CSIC and INTA and F.M. by grant
BMC2003-06957 from MEC (Spain).

%

\end{document}